\documentclass[conference]{IEEEtran}
\IEEEoverridecommandlockouts
\usepackage{cite}
\usepackage{amsmath,amssymb,amsfonts}
\usepackage{algorithmic}
\usepackage{graphicx}
\usepackage{textcomp}
\usepackage{xcolor}
\usepackage{subfigure}
\usepackage{multirow}
\usepackage{url}
\usepackage{flushend}
\def\BibTeX{{\rm B\kern-.05em{\sc i\kern-.025em b}\kern-.08em
    T\kern-.1667em\lower.7ex\hbox{E}\kern-.125emX}}
\begin{document}

\title{Open and Programmable 5G Network-in-a-Box: Technology Demonstration and Evaluation Results}

\author{
\IEEEauthorblockN{Adnan Aijaz, Ben Holden, and Fanyu Meng}
\IEEEauthorblockA{
\text{Bristol Research and Innovation Laboratory, Toshiba Europe Ltd., Bristol, United Kingdom}\\
firstname.lastname@toshiba-bril.com}
}    

\maketitle

\begin{abstract}
The fifth-generation (5G) mobile/cellular technology is a game changer for industrial systems. Private 5G deployments are promising to address the challenges faced by industrial networks. Programmability and open-source are two key aspects which bring unprecedented flexibility and customizability to private 5G networks. Recent regulatory initiatives are removing barriers for industrial stakeholders to deploy their own local 5G networks with dedicated equipment. To this end, this demonstration showcases an open and programmable \emph{5G network-in-a-box} solution for private deployments. The network-in-a-box provides an integrated solution, based on  open-source software stack and general-purpose hardware, for operation in 5G non-standalone (NSA) as well as 4G long-term evolution (LTE) modes. The demonstration also shows the capability of operation in different sub-6 GHz frequency bands, some of which are specifically available for private networks. Performance results, in terms of end-to-end latency and data rates, with a commercial off-the-shelf (COTS) 5G device are shown as well. 
\end{abstract}


\section{Introduction}
The industrial connectivity landscape is radically changing with the emergence of fifth-generation (5G) mobile/cellular technology which has unprecedented capabilities, particularly in terms of providing a unified wireless interface for diverse legacy and emerging applications \cite{3gpp.22.804}. Private deployments of 5G technology provide several benefits for industrial networks such as dedicated coverage, exclusive capacity, and intrinsic control \cite{pvt_5G}. Programmability in private 5G networks is important for fine-grained control and necessary customization for meeting the requirements of industry verticals. A prominent example is network slicing which has been shown to provide strict performance guarantees for industrial communication \cite{5G_demo}. On the other hand, a number of recent projects around software, architecture, and management aspects of 5G have adopted the open-source model \cite{open_5G_survey}. Such openness not only promotes innovation but also paves the way for developing white box 5G technology. 

A number of recent regulatory initiatives worldwide have opened shared licensed spectrum for private networks which empower industrial stakeholders to run their own local 4G/5G networks. Notable examples of shared licensed spectrum include the 3.7 - 3.8 GHz band in Germany, the 4.6 - 4.9 GHz band in Japan, and the 3.5 GHz band in the U.S. In private deployments, \emph{in-a-box} integrated solutions are desirable so that local networks can run in isolation while the enterprise retains control of on-premises equipment, without relying on external data centres. Such integrated in-a-box solutions also provide portable and rapidly deployable communication infrastructure for government and military services. 

\subsection{Demonstration Overview}
This  demonstration showcases an open and programmable \emph{5G network-in-a-box} solution for private deployments. It demonstrates a prototype implementation based on open-source software stack and general-purpose hardware for the radio access network (RAN) and the core network. It also demonstrates end-to-end connectivity and performance, with a commercial off-the-shelf (COTS) device, in different sub-6 GHz frequency bands, some of which are specifically opened for private networks.

\subsection{Novelty and Distinguishing Aspects}
To the best of our knowledge, this is one of the first technology demonstrations of 5G network-in-a-box solution operating in 5G non-standalone (NSA) and 4G long-term evolution (LTE) modes. Our core contribution is a prototype implementation of integrated in-a-box programmable solution for 5G NSA and 4G LTE operation in different frequency bands. We are also the first in terms of demonstrating successful end-to-end 5G NSA connectivity  with a \emph{COTS 5G dongle} and operation in recently-opened shared licensed spectrum for private networks in the U.K. \cite{ofcom_lic}. We also provide one of the first empirical evaluations of 5G NSA performance\footnote{A recent demonstration \cite{demo_NSA_OAI} shows connection procedure of 5G NSA, based on an open-source stack, with a \emph{COTS phone}. However, end-to-end performance of 5G NSA is largely unexplored.},  in terms of end-to-end latency and throughput based on an open-source software stack.

\begin{figure}
\centering
\includegraphics[scale=0.42]{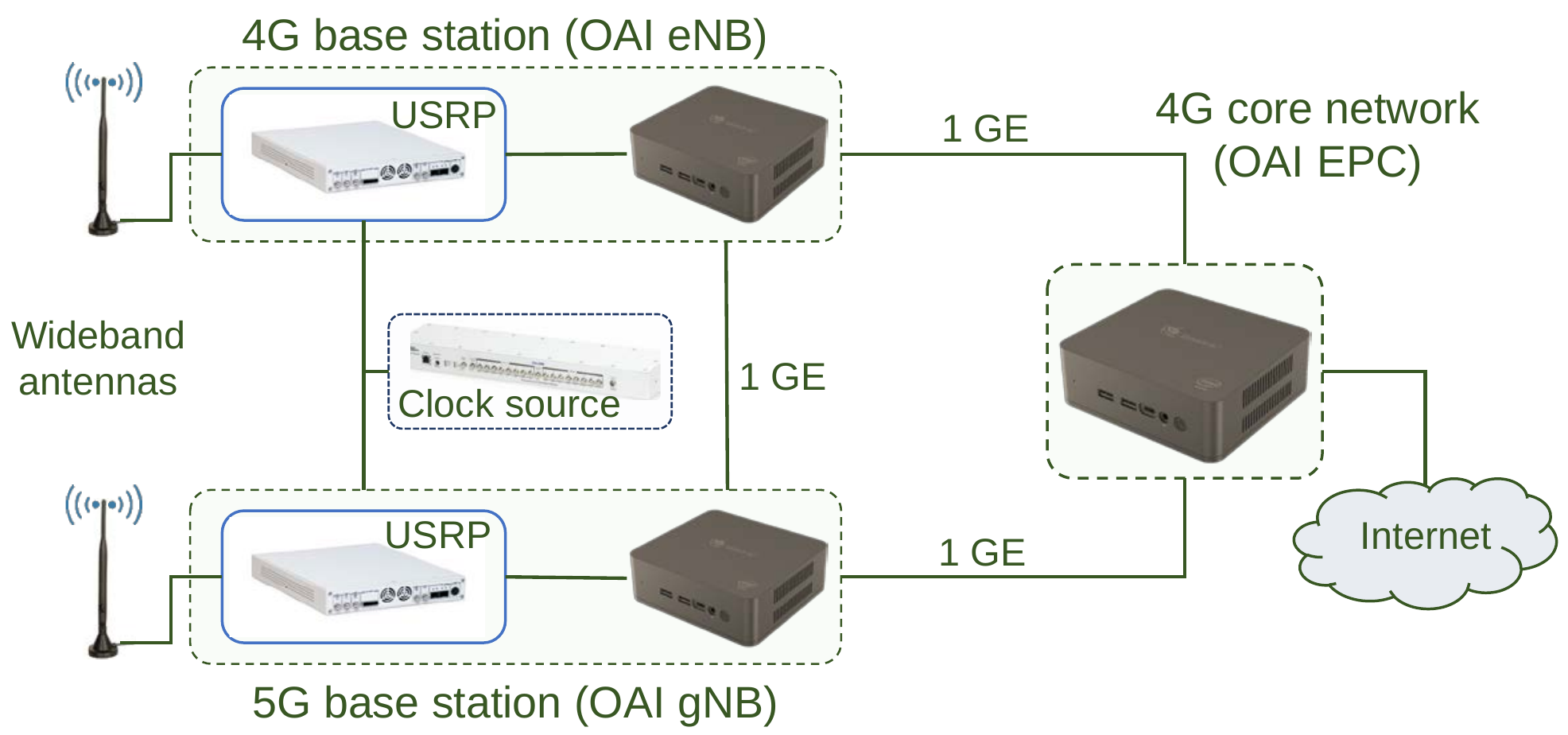}
\caption{Illustration of the 5G NSA system configuration.}
\label{config}
\end{figure}

\section{Design and Implementation Aspects}
\subsection{Network Configuration}
Our 5G network-in-a-box solution is based on the NSA configuration of 5G New Radio (NR), which is illustrated in Fig. \ref{config}. In the NSA mode, the 5G base station (gNB) co-exists with a 4G base station (eNB) and a 4G core network, also known as evolved packet core (EPC). The EPC consists of a mobility management entity (MME), a home subscriber server (HSS), and a serving and packet data network gateway (S-PGW). All control-plane functions are handled via the eNB whereas all user-plane functions are supported by the gNB. The two base stations are connected via a 1 Gigabit Ethernet (GE) interface. The eNB and the gNB are separately connected to the core network via 1 GE interfaces.

\begin{figure}
\centering
\includegraphics[scale=0.67]{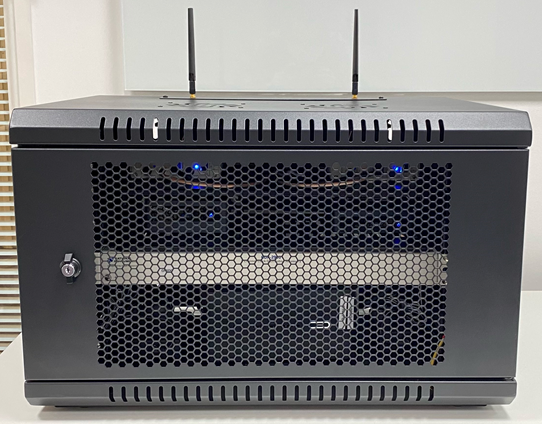}\\ 
\includegraphics[scale=0.72]{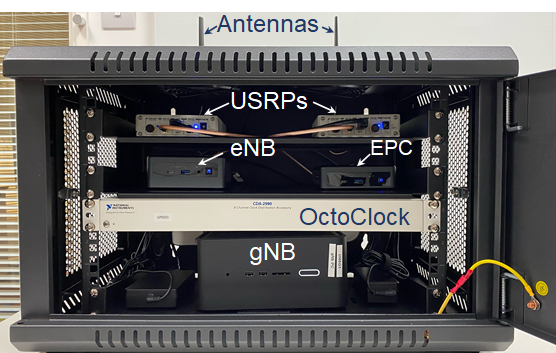}
\caption{Prototype of our 5G network-in-a-box solution.}
\label{box1}
\end{figure}

\subsection{Hardware and Software Specifications}
The eNB and the gNB front-ends are based on Ettus B210 software-defined radio (SDR) platforms, connected via USB 3.0 interfaces to respective servers. The eNB/gNB front-ends use wideband antennas. We use open-source software stack for the eNB, the gNB and the EPC based on OpenAirInterface (OAI) \cite{OAI_5G, OAI_5G_journal} specifications. The software stack\footnote{https://gitlab.eurecom.fr/oai/openairinterface5g/-/tree/develop}\footnote{https://github.com/OPENAIRINTERFACE/openair-epc-fed/tree/master} for each network entity is running on a dedicated mini PC kit. The eNB stack is running on an Intel NUC 10th Gen i7 mini PC with 16 GB RAM. The gNB stack is running on a higher specification Intel Core i9 Ghost Canyon Extreme NUC mini PC with 32 GB RAM. The EPC stack is running on an Intel NUC 10th Gen i5 mini PC with 16 GB RAM. We use Ettus OctoClock-G CDA-2990 for time synchronization of the eNB and the gNB USRPs. A prototype of our 5G network-in-a-box (450 x 600 x 368 mm) solution  is shown in Fig. \ref{box1}.

\subsection{COTS Devices}
\begin{figure}
\centering
\includegraphics[scale=0.66]{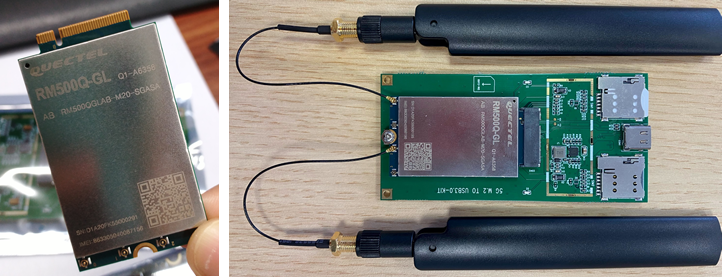}
\caption{Quectel's 5G module with M.2 to USB 3.0 adapter and antennas.}
\label{module}
\end{figure}

Our 5G network-in-a-box solution supports 4G/5G COTS devices. Our primary COTS device is a dongle which is based on Quectel RM500Q-GL module, an M.2 to USB adapter, and antennas connected to the module by UFL to SMA cables. This module (shown in Fig. \ref{module}) is compatible with 3GPP Release 15 and supports 5G NSA operation as well as 4G LTE operation in several sub-6 GHz frequency bands. 

In addition to the dongle, we have tested successful 5G NSA connection and end-to-end data transfer with a COTS phone (Samsung S10 5G).

\section{Operational Capabilities and Performance}
Our 5G network-in-a-box solution is capable of operating in 5G NSA and 4G LTE modes in different sub-6 GHz frequency bands. 
For this demonstration, we have conducted successful operation in the following bands.

\begin{itemize}
\item \emph{5G NSA Mode} -- band 78 (3.3 GHz -- 3.8 GHz) for gNB, with band 7 (2.6 GHz) and band 3 (1.8 GHz) as anchors. The anchor bands refer to those used by the 4G base station in NSA mode.

\item \emph{4G LTE Mode} -- band 3 (1.8 GHz) and band 40 (2.3 GHz). Note that dedicated spectrum within bands 3 and 40 has been been allocated for private networks in the U.K. \cite{ofcom_lic}.
\end{itemize}

In terms of duplexing mode, bands 78 and 40 support time division duplexing (TDD), whereas bands 7 and 3 support frequency division duplexing (FDD).  We have conducted end-to-end performance evaluation\footnote{In our demonstration, the eNB/gNB transmit at very low power as per the regulatory requirements for license-free operation in licensed bands.}, in terms of latency and throughput, with the 5G dongle.

\subsection{Operation in 5G NSA Mode}
Our 5G NSA implementation uses 40 MHz bandwidth and a sub-carrier spacing of 30 kHz. Fig. \ref{lat_res} shows the cumulative distribution function (CDF) of the round-trip latency between the 5G dongle and the EPC. The average round-trip latency of 5G NSA is 10.01 msec and a minimum value of 6.9 msec has been observed. 

We have conducted throughput evaluation based on TCP iPerf test between the 5G dongle and the EPC. Fig. \ref{thpt_res} shows the CDF of the downlink (DL) throughput for 5G NSA. Similar throughput performance has been observed with both anchor bands. The mean throughput in 5G NSA mode for anchor bands 3 and 7 is  22.14 Mbps and 22.21 Mbps, respectively. A maximum throughput of 30.7 Mbps has been observed for 5G NSA mode. Further throughput enhancement for 5G NSA is under investigation in the OAI community.



\subsection{Operation in 4G LTE Mode}
\begin{figure}
\centering
\includegraphics[scale=0.3]{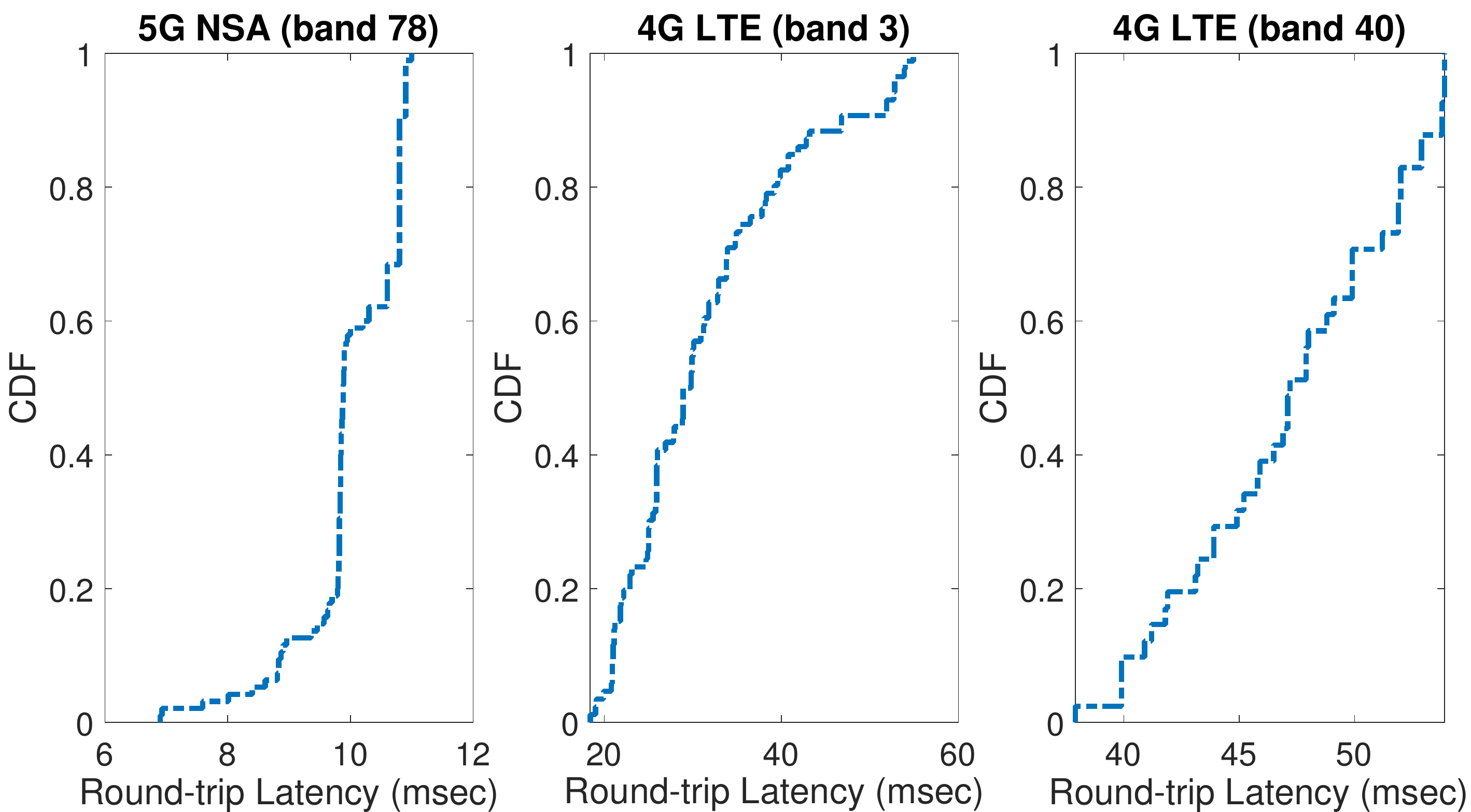}
\caption{Latency performance of 5G NSA and 4G LTE modes. }
\label{lat_res}
\end{figure}

We have evaluated the performance of 4G LTE mode in band 3 and band 40 with a bandwidth of 5 MHz for the former and 10 MHz for the latter. This bandwidth is approximately the same as that available for private networks in these bands \cite{ofcom_lic}. Fig. \ref{lat_res} shows the CDFs of round-trip latency performance (between the dongle and the EPC) of 4G LTE mode. Compared to 5G NSA mode, the achieved latency is significantly higher. The mean latency  in bands 3 and 40 is 31.45 msec and 47.31 msec, respectively. The higher latency of band 40 is mainly due to TDD mode of operation. 

\begin{figure}
\centering
\includegraphics[scale=0.3]{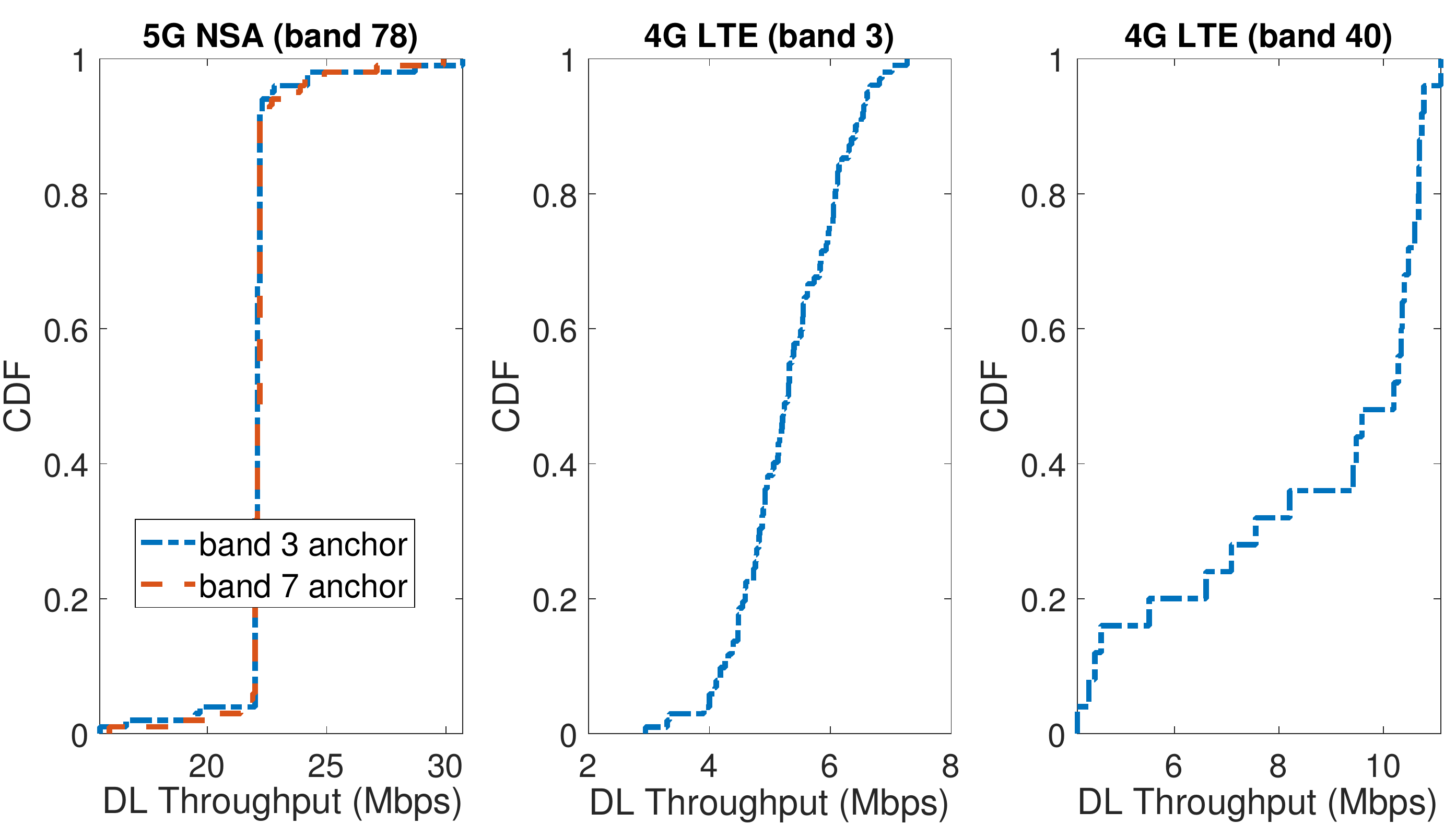}
\caption{Throughput performance of 5G NSA and 4G LTE modes. }
\label{thpt_res}
\end{figure}

Fig. \ref{thpt_res} shows the CDFs of DL throughput performance of 4G LTE mode based on TCP iperf test. The mean DL throughput in bands 3 and 40 is 5.31 Mbps and 8.73 Mbps, respectively. The better throughput of band 40 is due to higher bandwidth.


\section{Remarks}
This demonstration shows a 5G network-in-a-box solution for private deployments. Based on programmable hardware and open-source software stack, our integrated solution is capable of operating in  different sub-6 GHz frequency bands, in 5G NSA as well as 4G LTE modes. While the device ecosystem around private 5G is still in infancy, our demonstration shows a COTS dongle which can support 5G NSA (and 4G) operation in dedicated spectrum for private as well as public/commercial networks. Further, it shows the latency and throughput performance of the OAI-based 5G/4G stack in different frequency bands. Last but not least, it highlights the potential of general-purpose hardware and open-source software for developing white box 5G solutions.  Such solutions are becoming increasingly important as governments worldwide aim to diversify 5G supply chain with emphasis on open, resilient, secure, and innovative markets.

\section{Demonstration Setup}
Our demonstration setup consists of the 5G box and the 5G dongle (based on the Quectel RM500Q-GL module). We will demonstrate live connectivity and traffic exchange between the dongle and the network. The programmability of our solution will be demonstrated in the form of on-the-fly operation in 5G NSA and 4G modes in different frequency bands. We will also show live performance results for latency and throughput in different scenarios. 

\section{Acknowledgement}
This work has been supported by the UMBRELLA project (https://beta.southglos.gov.uk/umbrella-network/). The authors would also like to gratefully acknowledge the  contributions and support of OAI and Mosaic5G communities.

\bibliographystyle{IEEEtran}
\bibliography{demo}

\begin{thebibliography}{1}
\providecommand{\url}[1]{#1}
\csname url@samestyle\endcsname
\providecommand{\newblock}{\relax}
\providecommand{\bibinfo}[2]{#2}
\providecommand{\BIBentrySTDinterwordspacing}{\spaceskip=0pt\relax}
\providecommand{\BIBentryALTinterwordstretchfactor}{4}
\providecommand{\BIBentryALTinterwordspacing}{\spaceskip=\fontdimen2\font plus
\BIBentryALTinterwordstretchfactor\fontdimen3\font minus
  \fontdimen4\font\relax}
\providecommand{\BIBforeignlanguage}[2]{{%
\expandafter\ifx\csname l@#1\endcsname\relax
\typeout{** WARNING: IEEEtran.bst: No hyphenation pattern has been}%
\typeout{** loaded for the language `#1'. Using the pattern for}%
\typeout{** the default language instead.}%
\else
\language=\csname l@#1\endcsname
\fi
#2}}
\providecommand{\BIBdecl}{\relax}
\BIBdecl

\bibitem{3gpp.22.804}
\BIBentryALTinterwordspacing
3GPP, ``{Study on Communication for Automation in Vertical Domains},'' {3rd
  Generation Partnership Project (3GPP)}, TR {22.804}, Dec. 2017, {v1.0}.
  [Online]. Available:
  \url{http://www.3gpp.org/ftp//Specs/archive/22_series/22.804/}
\BIBentrySTDinterwordspacing

\bibitem{pvt_5G}
A.~{Aijaz}, ``{Private 5G: The Future of Industrial Wireless},'' \emph{IEEE
  Industrial Electronics Magazine}, vol.~14, no.~4, pp. 136--145, 2020.

\bibitem{5G_demo}
\BIBentryALTinterwordspacing
J.~Thota and A.~Aijaz, ``{Slicing-Enabled Private 4G/5G Network for Industrial
  Wireless Applications},'' in \emph{ACM International Conference on Mobile
  Computing and Networking (MobiCom)}, New York, NY, USA, 2020. [Online].
  Available: \url{https://doi.org/10.1145/3372224.3417325}
\BIBentrySTDinterwordspacing

\bibitem{open_5G_survey}
L.~Bonati \emph{et~al.}, ``{Open, Programmable, and Virtualized 5G Networks:
  State-of-the-Art and the Road Ahead},'' \emph{Computer Networks}, vol. 182,
  p. 107516, 2020.

\bibitem{ofcom_lic}
\BIBentryALTinterwordspacing
Ofcom, ``{Enabling Wireless Innovation through Local Licensing},'' {The Office
  of Communications}, Report, July 2019. [Online]. Available:
  \url{https://www.ofcom.org.uk/__data/assets/pdf_file/0033/157884/enabling-wireless-innovation-through-local-licensing.pdf}
\BIBentrySTDinterwordspacing

\bibitem{demo_NSA_OAI}
F.~Kaltenberger \emph{et~al.}, ``{OpenAirInterface 5G NSA System with COTS
  Phone},'' in \emph{ACM International Workshop on Wireless Network Testbeds,
  Experimental Evaluation \& Characterization (WiNTECH)}, New York, NY, USA,
  2020, p. 128–129.

\bibitem{OAI_5G}
N.~Nikaein \emph{et~al.}, ``{OpenAirInterface: A Flexible Platform for 5G
  Research},'' \emph{SIGCOMM Computer Communication Review}, vol.~44, no.~5, p.
  33–38, Oct. 2014.

\bibitem{OAI_5G_journal}
F.~Kaltenberger \emph{et~al.}, ``{OpenAirInterface: Democratizing Innovation in
  the 5G Era},'' \emph{Computer Networks}, vol. 176, pp. 107--284, 2020.

\end{thebibliography}
\end{document}